\documentclass[]{spie}  

\usepackage{amsmath,amsfonts,amssymb}
\usepackage{graphicx}
\usepackage{xcolor}
\usepackage[multi-part-units=single]{siunitx}
\usepackage{color}
\usepackage{bm}
\usepackage{url}
\usepackage[export]{adjustbox}
\usepackage{multirow}
\usepackage[colorlinks=true, allcolors=blue]{hyperref}

\title{Medulloblastoma Tumor Classification using Deep Transfer Learning with Multi-Scale EfficientNets}

\author[a]{Marcel Bengs}
\author[b,c,e]{Michael Bockmayr}
\author[b,c,d]{Ulrich Schüller}
\author[a]{Alexander Schlaefer}
\affil[a]{Institute of Medical Technology and Intelligent Systems, Hamburg University of Technology, Am
Schwarzenberg-Campus 3, Hamburg 21073, Germany;} 
\affil[b]{Department of Pediatric Hematology and Oncology, University Medical Center Hamburg-Eppendorf, Martinistraße 52, Hamburg 20246, Germany}
\affil[c]{Research Institute Children's Cancer Center Hamburg, Martinistraße 52, Hamburg 20251, Germany}
\affil[d]{Institute of Neuropathology, University Medical Center Hamburg-Eppendorf, Martinistraße 52, Hamburg 20246, Germany}
\affil[e]{Mildred Scheel Cancer Career Center HaTriCS4, University Medical Center Hamburg-Eppendorf, 20246 Hamburg, Germany }

\authorinfo{Further author information: (Send correspondence to Marcel Bengs)\\ Marcel Bengs: E-mail: marcel.bengs@tuhh.de\\
Alexander Schlaefer: E-mail: schlaefer@tuhh.de}

\begin{document} 
\maketitle 
\fontsize{10}{10}\selectfont
\begin{abstract}
Medulloblastoma (MB) is the most common malignant brain tumor in childhood. The diagnosis is generally based on the microscopic evaluation of histopathological tissue slides. However, visual-only assessment of histopathological patterns is a tedious and time-consuming task and is also affected by observer variability. Hence, automated MB tumor classification could assist pathologists by promoting consistency and robust quantification. Recently, convolutional neural networks (CNNs) have been proposed for this task, while transfer learning has shown promising results. In this work, we propose an end-to-end MB tumor classification and explore transfer learning with various input sizes and matching network dimensions. We focus on differentiating between the histological subtypes classic and desmoplastic/nodular. For this purpose, we systematically evaluate recently proposed EfficientNets, which uniformly scale all dimensions of a CNN. Using a data set with 161 cases, we demonstrate that pre-trained EfficientNets with larger input resolutions lead to significant performance improvements compared to commonly used pre-trained CNN architectures. Also, we highlight the importance of transfer learning, when using such large architectures. Overall, our best performing method achieves an F1-Score of 80.1$\%$. 

\end{abstract}

\keywords{Transfer learning, convolutional neural networks, digital pathology, histopathology, image analysis, medulloblastoma}


\section{INTRODUCTION}
Medulloblastoma (MB) is the most common malignant central nervous system tumor in childhood and constitutes a heterogeneous disease that can be divided in four main molecular subgroups, which are associated with different histopathological and clinical features \cite{northcott2019medulloblastoma}.
For establishing a diagnosis, pathologists assess microscopic histopathology slides or high-resolution images of digitized images and follow human-based decision rules. The histological subtypes of MB, which differ both in their appearance under the microscope and in their molecular tissue properties, impact the patient prognosis and the decision on the type of therapy \cite{northcott2019medulloblastoma}. According to the WHO \cite{louis20072007,northcott2019medulloblastoma} MB can be dived into four histological subtypes, classic type (CMB), desmoplastic/nodular type (DN), MB with extensive nodularity (MBEN) and large cell anaplastic MB (LCA). However, visual-only assessment of histopathology patterns requires expert knowledge and is a time-consuming task, while being affected by observer variability \cite{mosquera2014computer}. Hence, a decision support tool for pathologists that helps to classify the different histological subtypes would promote consistency and objective inter-observer agreement. \\
Recent studies have shown the feasibility of discriminating between different MB subtypes using feature extraction methods on image regions \cite{cruz2015method, lai2011texture, das2018study, das2020classification}. However, manual feature engineering is task-dependent and requires careful adaptation to the specific problem and data set \cite{phan2016transfer}. In contrast to that, the digitization of histology slides, e.g. the availability of high-resolution whole slide images (WSI) and the recent success of deep learning for various medical applications motivated end-to-end deep learning for digital pathology \cite{gertych2019convolutional, narayanan2019convolutional, bejnordi2017diagnostic}. Notably, convolutional neural networks (CNNs) outperform manual feature engineering on a variety of digital pathology tasks \cite{alom2019advanced, bejnordi2017diagnostic, phan2016transfer, kieffer2017convolutional, rachapudi2020improved}. However, training these models typically requires thousands of training examples, while annotated data is highly limited, especially for rare entities like pediatric brain tumors. A common approach to overcome limited data in digital pathology is transfer learning \cite{bejnordi2017diagnostic, phan2016transfer, kieffer2017convolutional}. Here, a network pre-trained on a different domain with a large annotated data set is transferred to the task at hand and optimized on few available annotated images. While different transfer learning strategies exist, a recent study demonstrates that fine-tuning a pre-trained network leads to the best results across various data sets  \cite{mormont2018comparison}. These networks, optimized on the natural image domain, mainly use a fixed input resolution, e.g. of $224\times224$ pixels \cite{Huang2017}, which induces a limiting factor for the extremely high-resolution images of digitized histology slides. \\ Recently, EfficientNets  \cite{tan2019efficientnet} have been proposed which uniformly scale all dimensions of a CNN, specifically input resolution, network depth, and width. This novel architecture design achieves top-performance on several benchmark data sets while using significantly fewer parameters than competitive architectures. A key advantage of pre-trained EfficientNets is transfer learning with larger input sizes and matching network dimensions. \\ In this work, we provide a systematic evaluation of pre-trained EfficientNets with different input resolutions for the challenging task of MB tumor classification. We use a data set with WSI from 161 different patients and focus on the task of differentiating between types CMB and DN. Overall, the contribution of this study is threefold:
(1) We systematically evaluate pre-trained EfficientNets for MB tumor classification and assess the impact of several different input resolutions.
(2) We reveal that pre-trained EfficientNets with an increased input size significantly outperform commonly used pre-trained networks such as VGG16, AlexNet, or ResNet50. (3) We demonstrate the importance of pre-training in combination with large-scale EfficientNets.

\begin{figure}
\centering
\includegraphics[width=1.0\textwidth]{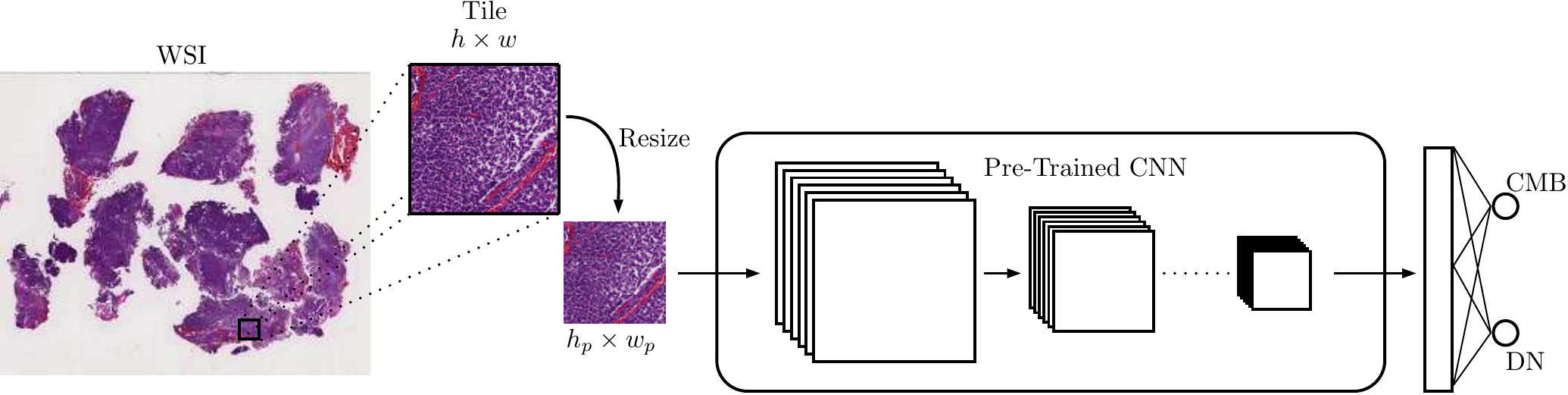}

\caption{Our models differentiate end-to-end the histological subtypes classic (CMB) and desmoplastic/nodular (DN). An image tile with dimensions $h \times w$ is cropped from the entire WSI. For classification, the tile is downsampled to the input resolution  $h_{p} \times w_{p}$ of a pre-trained network.}
\label{fig:example_images}
\end{figure}

\begin{figure}
\centering
\includegraphics[width=0.24\textwidth]{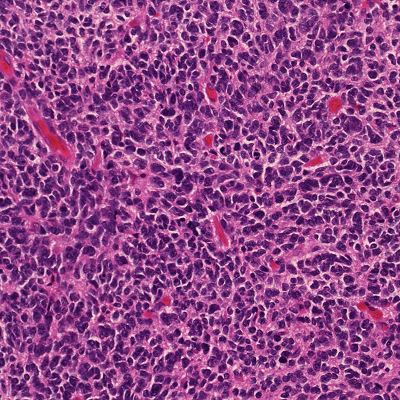}
\includegraphics[width=0.24\textwidth]{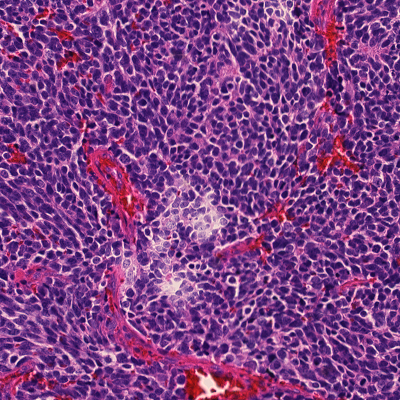} 
\hspace{0.25cm}
\includegraphics[width=0.24\textwidth]{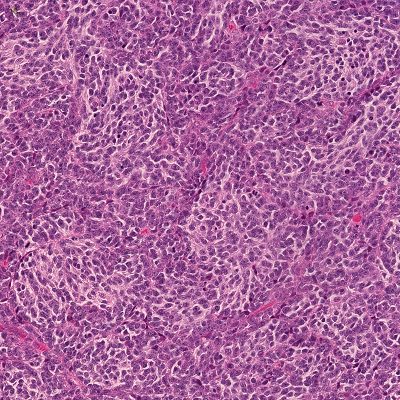}
\includegraphics[width=0.24\textwidth]{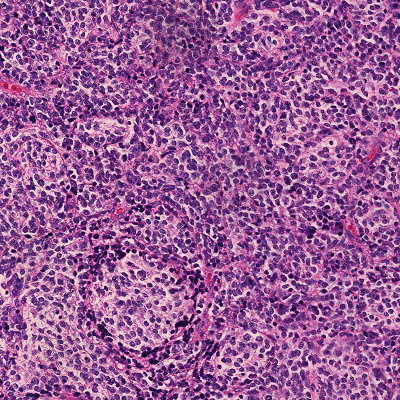}

\caption{Two example tiles for each of the histological subtypes classic type (CMB) and  desmoplastic/nodular type (DN); CMB is shown on the left side and DN is shown on the right side. }
\label{fig:example_images_v2}
\end{figure}

\section{Methods and Materials} 
\subsection{Data Set}
Our data set comes from 12 clinical sites in Germany and consists of haematoxylin \& eosin stained images from 161 patients (103 cases CMB and 58 DN) scanned at magnification 200x. Informed consent was obtained for all patients in accordance with local institutional guidelines. For data set labeling, cancerous regions in the WSI are labeled as CMB or DN by a neuropathologist. Of note, each WSI has one or more cancers regions. To generate our data set, we consider image tiles with a size of $2000\times2000$ pixels cropped from the manually annotated cancerous regions of the WSI. In total, we obtain 2769 image tiles from 161 patients with the corresponding label CMB or DN, example tiles for the two classes are shown in Figure \ref{fig:example_images_v2}.

\subsection{Deep Learning Methods} 
We consider various CNNs, pre-trained on ImageNet, with different input resolutions for MB tumor classification. Figure \ref{fig:example_images} describes our overall pipeline for classification. As a benchmark, we use established pre-trained CNNs, AlexNet \cite{krizhevsky2017imagenet}, VGG-16 \cite{simonyan2014very}, ResNet50 \cite{he2016deep}, and Densenet121 \cite{huang2017densely} that are widely used in digital pathology \cite{bejnordi2017diagnostic,phan2016transfer, kieffer2017convolutional, gertych2019convolutional, wang2016deep}. Next, we consider recently proposed EfficientNets \cite{tan2019efficientnet} which achieve state-of-the-art performance on several benchmark data sets, while having significantly fewer parameters than comparable models. The baseline of this architecture, called EfficientNet-B0, is optimized based on a multi-objective neural architecture search using the ImageNet data set. A special property of EfficientNet's concept is the compound scaling method, which uniformly scales network width, depth, and input resolution starting with the baseline EfficientNet-B0. The compound scaling is inspired by the idea that a larger input also needs a deeper and wider network, such that features can be learned effectively from the input. Using this compound scaling method, the authors of this network propose a set of EfficientNets with increasing scales of the network dimensions and input resolutions. In this work, we consider EfficientNet-B0 up to EfficientNet-B5, to systematically evaluate the impact of increased input resolution for MB tumor classification. Note, the corresponding input resolutions are shown in Table \ref{tab:All-networks-with metrics}. \\
For the evaluation of our models, we randomly split our data based on patients and consider 10-fold cross-validation. We equally split the data into a test and validation subset for each fold. Note, the subsets consist of five and two cases for type classic and desmoplastic/nodular, respectively. As the classification is highly imbalanced, we weight the loss of the individual classes inversely proportional to samples of each class. 
\\
Given an image tile with a resolution of $2000\times 2000$, we downsample the tile to the corresponding network input size. To counter stain variation, we employ extensive color augmentation during training, using brightness, contrast, saturation, and hue augmentation  \cite{tellez2019quantifying}. Also, we use random horizontal and vertical flipping of the images as an additional data augmentation strategy. Note, we evaluate our models based on tile classification performance. 

\section{Results} 
\begin{figure}
\centering
\includegraphics[width=0.60\textwidth]{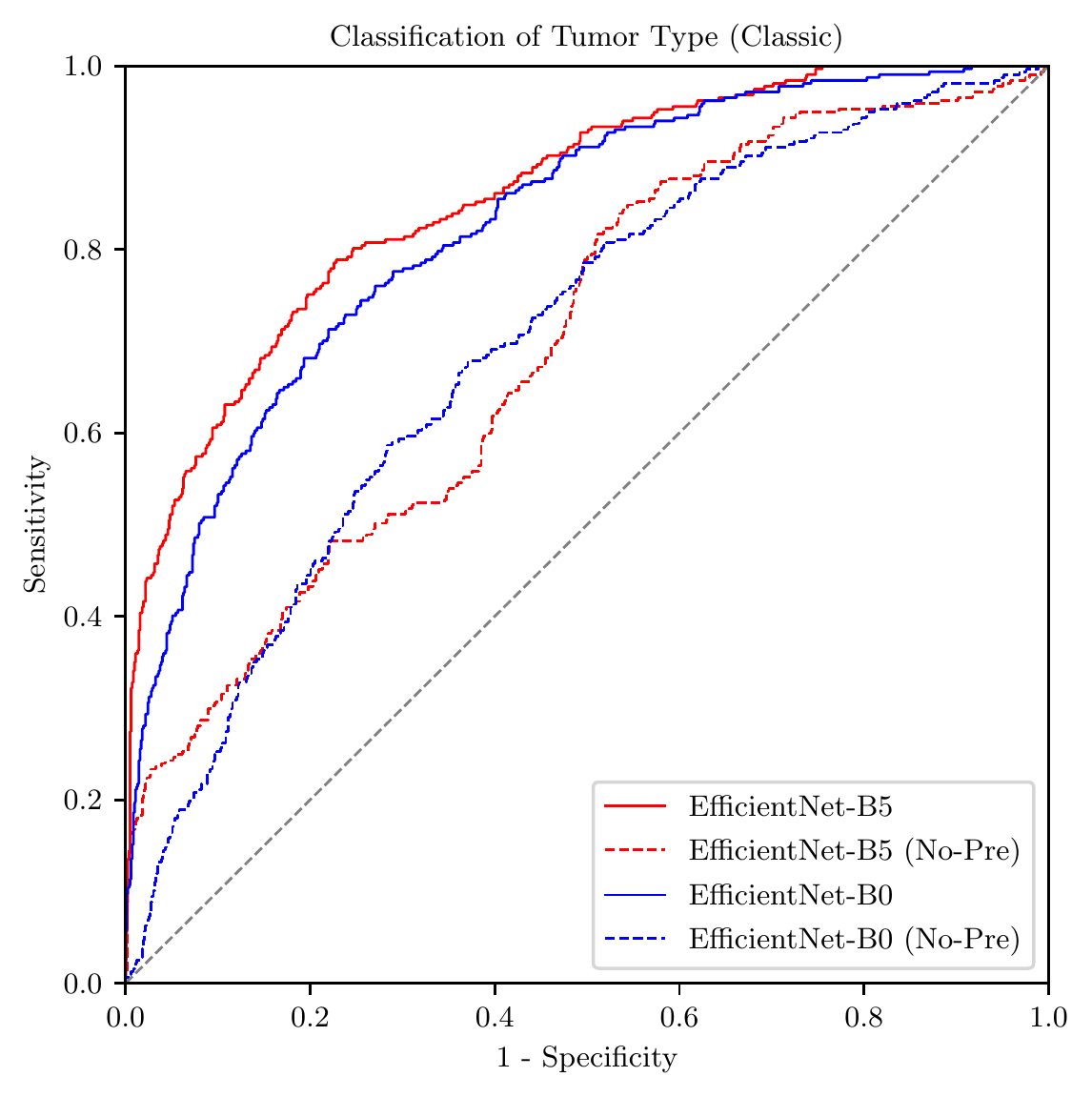}
\caption{ROC curve for pre-trained and non pre-trained EfficientNet-B0 and EfficientNet-B5, shown for the classification of MB tumor type.}
\label{fig:ROC}
\end{figure}

We report F1-Score, sensitivity, specificity and the area under the receiver operating curve (AUC) with $\SI{95}{\percent}$ confidence intervals (CI) using bias-corrected and accelerated bootstrapping with $n_{\mathit{CI}} = \num{10000}$ bootstrap samples in Table \ref{tab:All-networks-with metrics}. Note, we do not consider accuracy due to our imbalanced data set. For testing of significance, we use a permutation test with $n_{\mathit{P}} = \num{10000}$ samples and a significance level of $\alpha = 5\%$ \cite{efron1994introduction}. Overall, EfficientNet-B5 with the largest input resolution of $456\times456$ pixels performs best and AlexNet with an input resolution of $224\times224$ pixels performs worst. Moreover, we evaluate the impact of pre-training in Figure \ref{fig:ROC} using a receiver operating characteristic (ROC) curve. In this regard, pre-training leads to significant performance improvements, especially for EfficientNet-B5. Also, when no pre-training is used EfficientNet-B0 and EfficientNet-B5 perform similar.

\section{DISCUSSION} 

\begin{table}
 {\caption{Results for all experiments given in percent. Sensitivity and specificity are reported with respect to classifying an image as classic type. 95 \% CIs are provided in brackets. The best performing method is shown in bold.}  \label{tab:All-networks-with metrics}}%
\centering
  {\begin{tabular}{llllll}
  & Input Size &   AUC &  Sensitivity & Specificity &  F1-Score  \\ \hline
  
  AlexNet & $224\times 224$ & $69.3(66-73)$ &  $77.0(74-80)$ & $52.1(46-58)$ & $71(68-73)$ \\
  VGG-16 &$256 \times 256$  & $78.2(75-81)$ & $73.3(71-76)$  & $65.6(60-70)$ & $72.2(70-75)$  \\
  ResNet50  & $224 \times 224$ & $77.3(74-80)$ & $80.3(77-83)$  &  $60.2(55-65)$ & $75.1(73-78)$ \\
    Densenet121  & $224 \times 224$ & $78.2(75-81)$ & $79.5(77-82)$ & $61.8(57-67)$ & $75.2(73-78)$ \\

 EfficientNet-B0  & $224 \times 224$&$82.6(80-85)$ & $76.3(73-79)$ & $71.9(66-77)$ & $76.0(74-78)$ \\

  EfficientNet-B1 & $240 \times 240$& $82.5(80-85)$  & $78.7(76-81)$  & $68.7(64-74)$ & $76.6(74-79)$  \\
  
  EfficientNet-B2  & $260\times260$ & $83.6(81-86)$ & $80.0(76-82)$ & $71.6(66-76)$ &$77.7(75-80)$ \\
  
 EfficientNet-B3  & $300\times300$ & $85.5(82-87)$ & $81.3(78-84)$ & $72.2(67-77)$ & $79.3(77-82)$ \\
 
  EfficientNet-B4  & $380\times380$ &  $85.5(83-88)$ & $80.3(78-88)$ & $\pmb{75.3(70-80)}$ & $79.6(77-82)$ \\
  
   EfficientNet-B5  & $456\times456$ & $\pmb{85.8(83-88)}$ & $\pmb{83.7(81-86)}$ & $69.4(64-74)$ & $\pmb{80.1(78-82)}$ \\

\hline
  \end{tabular}}
\end{table}
We consider the task of MB tumor classification, which could assist pathologists by accelerating the visual assessment and reducing observer variability. While pre-trained CNNs with a small fixed input resolution are commonly used in digital pathology,  we evaluate pre-trained CNNs with different input resolutions and scales. We focus on differentiating between the histological subtypes DN and CMB using image tiles extracted from the WSIs. 
\\ 
Our results in Table \ref{tab:All-networks-with metrics} show steady performance improvements when novel architecture design principles and increased input resolutions are combined. Considering the different architectures shows that classical CNNs such as AlexNet and VGG-16 perform worse. Also, there is no significant difference between the F1-Score of ResNet, Densenet, and EfficientNet-B0 using a fixed input resolution of $224 \times 224$. This indicates that performance is limited due to the input resolution and simply increasing network width, depth, or architecture concept does not lead to significant performance gains. However, following the principle of EfficientNet and increasing network size and input resolution leads to significant performance improvements. In particular, using the largest input resolution of $465 \times 456 $ with EfficientNet-B5 performs best with an F1-Score of $80.1\%$ and significantly outperforms established pre-trained CNNs as well as EfficientNet-B0. This demonstrates that the concept of EfficientNet, e.g. uniformly scaling network width, depth, and input resolution allows for improved MB classification. Also, it has been shown that EfficientNet outperforms classical CNNs on a variety of transfer learning tasks \cite{tan2019efficientnet}, our results confirm this finding and highlight that the concept of EfficientNet also transfers well to MB classification, outperforming classical CNNs. \\
We evaluate the importance of pre-training in Figure \ref{fig:ROC}. Notably, when using no pre-training EfficientNet-B0 and EfficientNet-B5 perform similar and the larger input resolution does not lead to performance improvements. However, when pre-training is used, EfficientNet-B5 significantly outperforms EfficientNet-B0. This demonstrates the importance of pre-training when using such large-scale architectures, which may suffer from an increased risk of overfitting the training data, especially when training data is highly limited such as in the case of MB classification. Note, EfficientNet is pre-trained on natural images from the ImageNet challenge. Thus, pre-training EfficientNets on large-scale histopathology image data sets even from different organs or different diagnostics tasks might allow for further performance improvements, taking into account the findings of a previous study on MB classification and transfer learning \cite{cruz2015method}.  Moreover, our study focuses on classifying image tiles pre-extracted from WSIs into the two classes CMB or DN. Hence, the classification of the entire WSI remains an open challenge that could be addressed by combining our findings with previous works on WSI classification, where a CNN, such as ResNet50, VGG-16 or inception-v3 is applied to the WSI in a sliding window fashion \cite{iizuka2020deep,vsaric2019cnn}. In this regard, using a large-scale EfficientNet instead of classical CNNs for tile feature extraction might be a promising approach.  
Overall, our results provide a good starting point to further improve tile classification performance, by using large scale pre-trained EfficientNets.  \\

\section{CONCLUSION} 
In this work, we address the task of MB tumor classification and highlight the advantage of larger input resolutions and novel architecture design principles combined with transfer learning. In this context, we provide a comprehensive study on different network architectures using transfer learning or training from scratch. Results of our study demonstrate significant performance improvements by using large scale pre-trained EfficientNet compared to compared to commonly used pre-trained CNN architectures. Future work could focus on classifying all subtypes of MB using a larger data set. Moreover, our findings could be extended to other classification problems. \\ \\
\textbf{Acknowledgments.} This work was partially supported by the Hamburg University of Technology i$^{3}$ initiative. 

\bibliography{report}   
\bibliographystyle{spiebib}   

\end{document}